\documentstyle[aps]{revtex}
\input{epsf}
\begin{document}
\title{Stability under radiation reaction
of circular equatorial orbits around Kerr black holes}
\author{Daniel Kennefick}
\address{SOCAS, 50 Park Place, University of Wales, Cardiff, Wales
CF1 3AT}
\maketitle
\begin{abstract}
We examine the
evolution, under gravitational radiation reaction,
of slightly eccentric equatorial orbits of point particles around
Kerr black holes. Our method involves numerical integration
of the Sasaki-Nakamura equation. 
It is discovered that such orbits decrease in eccentricity 
throughout most of the inspiral, until shortly before the 
innermost stable circular orbit (ISCO), when a critical radius
$r_{\text{crit}}$ is reached beyond which the inspiralling orbits increase
in eccentricity. It is shown that the number of orbits remaining
in this last (eccentricity increasing) phase of the inspiral is
an order of magnitude less for prograde orbits around rapidly
spinning black holes than for retrograde orbits. In the extreme limit
of a Kerr black hole with spin parameter $a=1$,
this critical radius descends into the ``throat'' of the black
hole.
\end{abstract}
\section{Introduction}

Gravitational waves emitted by solar-mass-size compact bodies
orbiting massive ($10^6 M_\odot$ and greater) black holes (and
spiralling towards them as they lose energy and angular momentum
to the emitted radiation) are a favoured source for gravitational
wave detectors sensitive to low-frequency radiation, such as
proposed space-based detectors like the Laser Interferometer
Space Antenna (LISA) \cite{LISA4}. 
Systems of this type lend themselves to theoretical analysis
via perturbation theory, because of the extreme mass ratio between
the two bodies. In recent years, the Teukolsky perturbation formalism
for black holes has been employed successfully to describe orbital
decay of small bodies orbiting a large Schwarzschild (i.e. non-rotating)
black hole \cite{III,gen,Jap}, using a mixture of analytic and numerical
results (for an informative 
overview of much of the purely analytic work to date, see
Mino et al. \cite{MJap}). 
One result of this work has been
to modify the long-standing result \cite{P4} that, under radiation
reaction, orbits tend to become more circular as they slowly decay.
In fact, inside a critical radius, which is $r_{\text{crit}}=6.6792M$
for nearly circular orbits in the Schwarzschild geometry, non-circular
orbits tend to become more, rather than less, 
eccentric \cite{III}. 
Although a precisely circular
orbit would remain circular inside the critical radius, its circularity
is no   
longer stable to small perturbations away from precise circularity  
as the orbital decay continues.

Despite their intrinsic interest, these results may prove of limited
usefulness for any future low-frequency gravitational wave detectors,
since there is no reason to expect that large black holes should
typically have no spin at all. Just the opposite (that they should
exhibit strong rotation) is perhaps more to be expected
\cite{Bardeen}. Therefore it
is of great interest to extend this type of analysis to the case
of rotating, or Kerr black holes. This presents no difficulty for
the Teukolsky perturbation formalism itself, which was developed 
for the Kerr metric, but a problem does arise in dealing with
an additional constant of the motion which governs orbits around
spinning black holes. Unlike the energy and angular momentum, whose
flux can easily be determined from the waves far from the source,
until very recently there was 
no clear understanding of how to calculate the amount of ``Carter
constant'' carried away by the emitted radiation. In spite of this, it
has been shown recently, for general orbits in Kerr, that circular
orbits (defined as orbits of constant Boyer-Lindquist radius, and 
sometimes referred to as ``quasi-circular'') remain circular under
radiation reaction \cite{KO,Fin,Mino4}. While progress continues 
in developing techniques for dealing 
with general orbits in Kerr \cite{QW,MST,Amos}, it now seems worthwhile to
investigate the case of nearly-circular, equatorial orbits around
rotating black holes \cite{tag}. Equatorial orbits in the Kerr spacetime, like
orbits in Schwarzschild, can be said to have zero ``Carter constant'',
which remains unchanged during orbital decay. Looking at these orbits
can tell us if the behaviour previously observed for
slightly-eccentric orbits in Schwarzschild is also seen in the Kerr
metric for all values of the Kerr spin parameter $0\leq a\leq 1$.

It is shown in this paper that, for equatorial orbits, it is generally
true that a critical radius, $r_{\text{crit}}$ 
exists beyond which slightly eccentric
orbits become less circular due to radiation reaction,  and that
this radius is encountered close to the radius of the innermost
stable circular orbit (ISCO). This is best illustrated by
examining the behavior of the parameter $c=r_0 \dot{e}/
e \dot{r_0}$, where $e$ is the orbital eccentricity, and
$r_0$ the mean radius, and an overdot indicates differentiation by time.
This parameter is negative for orbits evolving with
increasing eccentricities,
and positive for decreasing eccentricity. Near the ISCO one can
show, as in Sec. 9 below, that $c$ diverges towards negative
infinity near the ISCO, for nearly all values of $a$. There is an 
apparent exception to this behaviour in the
limiting case of a maximally rotating Kerr black hole with $a=M$.
In that case, the horizon and the ISCO are both located at $r=M$
in Boyer-Lindquist co-ordinates, although they are still separate 
in terms of proper radial distance. As one approaches 
$r=M$, for the case of a prograde orbit around an extreme Kerr 
black hole, $c$ is both postive and finite, approaching the limit
of $3/2$ at $r=M$.
Not surprisingly therefore, for prograde orbits around black holes with very
large $a>.99M$, the transition to eccentricity-increasing inspiral
takes place only shortly before the onset of dynamical instability
at the ISCO in terms of the Boyer-Lindquist radial co-ordinate. 
The number of orbits remaining
at this point is an order of magnitude fewer for such cases than
it is for retrograde orbits in the same geometry. 

Since the radius
of the ISCO is much smaller for prograde than for retrograde orbits
(with $a= M$, $r_{\text{ISCO}}= M$ for prograde orbits and
$r_{\text{ISCO}}=9 M$ for retrograde orbits), the critical radius
is also much smaller for prograde orbits. These results demonstrate
that the onset of ``back reaction instability'' for circular orbits
precedes, and is intimately connected with, the onset of dynamical
instability signified by the ISCO. It seems reasonable to conjecture
that the alteration in the shape of the radial protential as  
the ISCO approaches,
at which point the minimum of the effective potential vanishes,
is reponsible for the gain in eccentricity. 

The organisation of the paper is as follows. In section 2, the orbital
equations for geodesic motion (i.e. not including radiation reaction)
are solved analytically 
for slightly eccentric, equatorial orbits. In section 3
the Tuekolsky perturbation formalism is described, and section 4
shows how to calculate the fluxes of energy and angular momentum
carried away from the system using this formalism. In section 5 
the Sasaki-Nakamura equation, which is actually solved rather than
the Tuekolsky radial equation for numerical reasons, is presented.
In section 6 the Teukolsky source function is calculated for a
perturbing particle following the orbits of section 2, and the
results of both of these sections come together in section 7 to
yield the rate of change of orbital eccentricity due to radiation
damping. This orbital evolution is described under the assumption
of adiabaticity (that the orbital evolution is much slower than the
orbital period), which introduces constraints which 
are discussed in section 8. 
Finally, in section 9, the analytic and numerical results are
presented, followed by a discussion of their significance in section
10. A guide to the essential points of the argument is given at
the end of section 7.

\section{Description of the orbit}

Since the perturbation of the Kerr metric producing the gravitational
waves
is that of a small particle orbiting
the black hole, it will be necessary to solve the orbital 
equations for a particle in orbit around a rotating black hole.
We require
expressions for $r(t)$, $\phi(t)$ and 
$\theta(t)$  to describe the orbit in Boyer-Lindquist co-ordinates.
Since we restrict ourselves to equatorial orbits, the solution for
the $\theta$ motion is trivial, $\theta=\pi/2$ is a constant
throughout. The equatorial 
orbital equations for a particle in the Kerr spacetime,
in these co-ordinates (leaving aside the trivial
$d\theta/d\tau=0$), are well known \cite{MTW4} 
\begin{eqnarray}
\mu \Sigma^2 dr/d\tau&=&[(E(r^2+a^2)-a L_z)^2
-\Delta(\mu^2 r^2+(L_z-a E)^2)]^{1\over2}\equiv \sqrt{R} \\
\mu \Sigma^2 d\phi/d\tau&=&-(a E-L_z/\sin^2 \theta)
+(a/\Delta)(E(r^2+a^2)-a L_z)\equiv \Phi \\
\mu \Sigma^2 dt/d\tau&=&-a(a E \sin^2 \theta
-L_z)+{(r^2+a^2)\over \Delta}(E(r^2+a^2)-a L_z)\equiv T,
\label{geo}
\end{eqnarray}
where $\tau$ is proper time,
$\Sigma=r^2+a^2\cos^2\theta$, $\Delta=r^2-2 M r+a^2$, and the black
hole's spin parameter $a$ is defined for convenience as
$a=\vec{J} \cdot \hat{L}/M$,
with $\vec{J}$ the spin angular momentum vector of
the black hole, and $\hat{L}$ a unit vector pointing in the
direction of the particle's orbital angular momentum vector.
For prograde orbits (in which the particle
orbits in the same sense as the black hole's spin) $a$ is positive,
and for retrograde orbits (in which the particle rotates in the opposite
sense to the hole), $a$ is negative. Recall that we restrict attention
to equatorial orbits only, so that $\vec{J}$ and $\hat{L}$ are either
parallel or anti-parallel. 
It is a condition of the
perturbation scheme that $\mu/M \ll 1$, where $M$ is the mass of
the central black hole
and $\mu$ the mass of the orbiting particle.
Finally, $E$ and $L_z$ are the particle's orbital energy
and angular momentum, respectively.

We now consider slightly eccentric orbits, and 
define a mean radius $r_0$, so that $\partial(R/r_0^4)/\partial r
|_{r=r_0}=0$. The eccentricity $e$ is defined so that $R(r=r_0(1+e))=0$.
These definitions are chosen so that as $e\rightarrow 0$, $r_0$ reduces
to the constant radius of a circular orbit, and so that $e$ corresponds,
when $e\ll1$ and in the appropriate limits,
to definitions of the eccentricity of an orbit commonly used in the 
Schwarzschild geometry and in Newtonian mechanics
\cite{III}. These defining equations for $r_0$ and $e$ permit us
to write the orbital energy and angular momentum in terms of these
two quantities. Since we assume throughout that $e$ is a small
quantity, it is convenient to expand $E$ and $L_z$
in terms of it, 
\begin{eqnarray}
E(r_0,e)&=&E_0(r_0)+e E_1(r_0)+e^2 E_2(r_0)+e^3 E_3(r_0)+O(e^4) \\
L_z(r_0,e)&=&L_0(r_0)+e L_1(r_0)+r^2 L_2(r_0)+e^3 L_3(r_0) +O(e^4).
\label{EL}
\end{eqnarray}
Using our two equations in $r_0$ and $e$, it is easy to show that
\begin{eqnarray}
E_0&=&\mu {1-2 v^2+q v^3\over(1-3v^2+2q v^3)^{1\over2}} \\
E_1&=& 0 \\
E_2&=&\mu {v^2(1-3v^2+q v^3+q^2v^4)(1-6v^2+8q v^3-3q^2v^4)\over 
2(1-3v^2+2q v^3)^{3\over2}(1-2v^2+q^2v^4)} \\
E_3&=&-\mu {v^2(1-3v^2+q v^3+q^2v^4)(1-7v^2+10q v^3-4q^2v^4)\over
(1-3v^2+2q v^3)^{3\over2}(1-2v^2+q^2v^4)} \\
L_0&=&\mu {r_0 v(1-2q v^3+q^2v^4)\over(1-3v^2+2q v^3)^{1\over2}} \\
L_1&=&0 \\
L_2&=&\mu {q r_0 v^5(q-3v+q v^2+q^2v^3)(1-6v^2+8q v^3-3q^2v^4)\over
2(1-3v^2+2q v^3)^{3\over2}(1-2v^2+q^2v^4)} \\
L_3&=&-\mu {q r_0 v^5(q-3v+q v^2+q^2v^3)(1-7v^2+10q v^3-4q^2v^4)\over
(1-3v^2+2q v^3)^{3\over2}(1-2v^2+q^2v^4)}. \label{lots}
\end{eqnarray}
Here $v=\sqrt{M/r_0}$ and $q=a/M$. These results, up to order
$e^2$ are given in Ref.\ \cite{tag}.

We wish to write the change in the eccentricity brought about by
the loss of orbital angular momentum and energy, in terms of the rates
of loss of those two quantities. Since we have $E$ and $L_z$ as 
functions of $r_0$ and $e$, we use the chain rule for differentiation
to write
\begin{eqnarray}
\dot{E}&=&-d E_{GW}/dt={\partial E\over \partial e}\dot{e}
+{\partial E\over \partial r_0}\dot{r}_0 \\
\dot{L}_z&=&-d L_{GW}/dt={\partial L_z\over\partial e}\dot{e}
+{\partial L_z\over\partial r_0}\dot{r}_0, \label{ELdot}
\end{eqnarray}
where $dE_{GW}/dt$ and $dL_{GW}/dt$ are the total energy
and angular momentum carried towards infinity and the black hole horizon
per unit time
by the gravitational waves, averaged over several wavelengths. We
will write these quantities also in terms of $e$ and $r_0$,
\begin{eqnarray}
{d E_{GW}\over dt}&=&\dot{E}_0+e \dot{E}_1+e^2 \dot{E}_2+O(e^3) \\
{d L_{GW}\over dt}&=&\dot{L}_0+e \dot{L}_1+e^2 \dot{L}_2+O(e^3).
\label{ELGW}
\end{eqnarray}
As we shall find later, $\dot{E}_1=\dot{L}_1=0$. Eliminating $r_0$
from Eq.\ (\ref{ELdot}), we derive
\begin{equation}
\dot{e}=[-{dE_{GW}\over dt}L_z^\prime+{dL_{GW}\over dt}E^\prime]/
[{\partial E\over\partial e} L_z^\prime-{\partial L_z\over\partial e}
E^\prime], \label{edot}
\end{equation}
where $\prime \equiv \partial/\partial r_0$. 

Substiting Eqs.\ (\ref{ELGW}) and (\ref{EL}) into Eq.\ (\ref{edot}),
we find, keeping terms up to order $e^2$,
\begin{equation}
\dot{e}={-L_0^\prime(\dot{E}_0-{E_0^\prime\over L_0^\prime}\dot{L}_0)-
e^2 L_0^\prime (\dot{E}_2-{E_0^\prime\over L_0^\prime}\dot{L}_2)-
e^2 L_2^\prime (\dot{E}_0-{E_2^\prime\over L_2^\prime}\dot{L}_0) \over
2 e (E_2 L_0^\prime-L_2 E_0^\prime)}.
\end{equation}
Now, from Eqs.\ (\ref{lots}), we see that
\begin{equation}
{E_0^\prime\over L_0^\prime}={\sqrt{M}\over r_0^{3\over2}+a\sqrt{M}}=
\Omega,
\end{equation}
where $\Omega$ is the angular frequency of a circular orbit of
radius $r_0$. It follows from an interesting (and quite general
\cite{Cart}) characteristic of
circular orbits, and will be shown later in this case that, the circular (i.e.
zeroth order in the eccentricity) rates
of loss of energy and angular momentum are related by
\begin{equation}
\dot{E}_0=\Omega \dot{L}_0 \label{cc}.
\end{equation}
Therefore
\begin{equation}
\mu \dot{e}=-e j(v)[g(v) \dot{E}_0
+\dot{E}_2-\Omega \dot{L}_2], \label{edotf}
\end{equation}
where 
\begin{equation}
j(v)={\mu\over E_2-\Omega L_2}={(1+q v^3)(1-2v^2+q^2v^4)
(1-3v^2+2q v^3)^{1\over2}\over v^2(1-6v^2+8q v^3-3q^2v^4)}
\end{equation}
and
\begin{equation}
g(v)={L_2^\prime\over L_0^\prime}-{E_2^\prime\over E_0^\prime}=
{{\cal G}(v)
\over 2(1+q v^3)(1-6v^2+8q v^3-3q^2v^4)(1-2v^2+q^2v^4)^2},
\end{equation}
where 
\begin{eqnarray}
{\cal G}(v)=
2&-&27v^2+72v^4-36v^6+38q v^3-17q^2v^4-144q v^5+86q^2v^6 \nonumber \\ 
&+& 4q^3v^7
+72q v^7-12q^4v^8-36q^2v^8-23q^4v^{10}+30q^5v^{11}
\nonumber \\ &-& 9q^6v^{12}
\end{eqnarray}
Since $\dot{e}$ is proportional to $e$ in this equation, it is plain
that a precisely circular orbit (one with $e=0$), will remain circular under
radiation reaction, provided that we can indeed show that
$\dot{E}_0=\Omega \dot{L}_0$ and $\dot{E}_1=\dot{L}_1=0$. It is also
plain that the question of the stability of an orbit's circularity 
will be determined by the sign of 
Eq.\ (\ref{edotf}), which requires us to calculate the loss of orbital energy
and angular momentum up to second order in $e$.

Similarly we can solve for $\dot{r}_0$, the rate of change of the
orbital radius, which tells us that to leading order $\dot{r}_0=
-\dot{E}_0/E^\prime_0$, which implies that
\begin{equation}
\mu \dot{r}_0/r_0=-{2(1-3v^2+2q v^3)^{3/2}\over v^2(1-6v^2+8q v^3-3q^2v^4)}
\dot{E}_0. \label{rdot}
\end{equation}

With this in hand it is possible to proceed to the solution of the
geodesic equations [Eqs.\ (\ref{geo})]. We expand $r(t)$ about the mean
radius $r_0$ in terms of the small eccentricity $e$, so that
\begin{equation}
r(t)=r_0[1+e r_1(t)+e^2 r_2(t)+O(e^3)]. \label{R}
\end{equation}
Making use of the expansions of $E$, $L_z$ and $r(t)$ in terms of
$e$, we expand out the equation $(dr/dt)^2=R/T^2$, and collect
terms of order $e^2$ and $e^3$ (note that the $e^3$ term in $r(t)$
does not contribute until $O(e^4)$ in $R/T^2$), giving us two differential
equations,
\begin{equation}
(d r_1/dt)^2=\Omega_r^2(1-r_1^2),
\end{equation}
where we define a radial frequency, 
\begin{equation}
\Omega_r=\Omega (1-6v^2+8q v^3-3q^2v^4)^{1\over2}
\end{equation}
and
\begin{equation}
{1\over\Omega_r^2}{d r_1\over dt}{d r_2\over dt}+r_1 r_2=
f_1(v)+f_2(v) r_1+f_3(v) r_1^3,
\end{equation}
where
\begin{eqnarray}
f_1(v)&=&-{1-7v^2+10q v^3-4q^2v^4\over 1-6v^2+8q v^3-3q^2v^4} \\
f_2(v)&=&{2v^2(1-2q v^3+q^2v^4)\over (1+q v^3)(1-2v^2+q^2v^4)} \\
f_3(v)&=&{{\cal F}_3(v)
\over (1+q v^3)(1-2v^2+q^2v^4)(1-6v^2+8q v^3-3q^2v^4)},
\end{eqnarray}
and
\begin{eqnarray}
{\cal F}_3(v)=
1&-&11v^2+26v^4+11q v^3-3q^2v^4-41q v^5+15q^2v^6 \nonumber \\&-&10q
v^7+7q^3v^7+24q^2v^8-4q^4v^8-27q^3v^9+16q^4v^{10}
\nonumber \\ &-& 4q^5v^{11}.
\end{eqnarray}
Integrating these equations in order, we find,
\begin{eqnarray}
r_1(t)&=&\cos (\Omega_r t) \\
r_2(t)&=&-f_1(v)(1-\cos(\Omega_r t))+{1\over2}f_3(v)(1-\cos(2\Omega_r t)).
\end{eqnarray}

It remains to solve for the $\phi$-motion. Again we expand out the
geodesic equation $d\phi/dt=\Phi/T$, integration of which yields
\begin{equation}
\phi(t)=\Omega_\phi t -e p(v) \sin (\Omega_r t)+O(e^2), \label{Phi}
\end{equation}
where 
\begin{equation}
p(v)={2(1-3v^2+2q v^3)\over[(1+q v^3)(1-2v^2+q^2v^4)(1-6v^2+8q v^3
-3q^2v^4)^{1/2}]}
\end{equation}
and
\begin{eqnarray}
\Omega_\phi&=&\Omega\Bigl[1 \nonumber \\ &-&
{3(1-11v^2+24v^4+13q v^3-4q^2v^4-46q v^5+25q^2v^6
+q^3v^7-3q^4v^8)\over2(1+q v^3)(1-2v^2+q^2v^4)(1-6v^2+8q v^3-3q^2v^4)}
e^2 \nonumber \\ &+& O(e^3)\Bigr] 
\nonumber \\ &\equiv&\Omega[1-\Delta \Omega e^2+O(e^3)]
\end{eqnarray}
is the azimuthal angular frequency.
The $O(e^2)$ part of $\phi(t)$ which is proportional to $\sin (\Omega_r
t)$ is not given, as neither it nor the $O(e^2)$ part of $r(t)$
contribute to the final result for $\dot{e}$, for reasons which will
become clear later. Only the $O(e^2)$ part of $\Omega_\phi$ 
(i.e. $\Delta \Omega$) is
required, although it is necessary to know $r_2(t)$ to derive $\Delta
\Omega$.

\section{The Teukolsky formalism}

We employ a scheme previously used in the Schwarzschild case to
investigate the evolution of 
slightly eccentric orbits under radiation reaction
\cite{III}. This scheme is 
based on the Teukolsky formalism for perturbations of the Kerr
metric. In this formalism one can decompose the Weyl scalar $\psi_4$
(which describes gravitational wave fluxes near infinity for such
a system) as follows,
\begin{equation}
\psi_4={1\over (r-i a \cos\theta)^4}\int^{+\infty}_{-\infty}
\sum_{lm} R_{lm\omega}(r) \mbox{}_{-2} S_{lm}^{a \omega}(\theta) e^{i m \phi}
e^{-i \omega t} d\omega, \label{weyl}
\end{equation}
where $\mbox{}_{-2}S_{lm}^{a \omega}$ is the spheroidal harmonic function of
spin weight $s=-2$. The normalization used here for these functions
is $\int^\pi_0 |\mbox{}_{-2} S_{lm}^{a \omega}(\theta)|^2 \sin\theta d\theta=
1/ 2\pi$. The radial function $R_{lm\omega}(r)$ obeys the
Teukolsky equation,
\begin{equation}
\Delta^2{d\over dr}\Bigl({1\over \Delta}{dR_{lm\omega}\over dr}\Bigr)-
V(r) R_{lm\omega}(r)=T_{lm\omega}(r) \label{Tr},
\end{equation}
where $T_{lm\omega}$ is the Teukolsky source function, to be evaluated
below. The Teukolsky potential is defined by
\begin{equation}
V(r)=-{K^2+4 i(r-M)K\over \Delta}+8 i \omega r +\lambda,
\end{equation}
where $K=(r^2+a^2)\omega-m a$ and $\lambda$ is the eigenvalue
associated with the appropriate spheroidal harmonic 
$\mbox{}_{-2} S^{a\omega}_{lm}$.

We can define two solutions to the homogeneous Teukolsky
equation, $R^H_{lm\omega}(r)$
and $R^\infty_{lm\omega}(r)$, with the following boundary conditions,
\begin{eqnarray}
R^H_{lm\omega} &\sim& \Delta^2 e^{i k r^*}, \text{as} \quad 
r\rightarrow r_+ \label{sola} \\
R^H_{lm\omega} &\sim& r^3 B^{\text{out}}_{lm\omega} e^{i\omega r^*}+
{1\over r}B^{\text{in}}_{lm\omega} e^{-i\omega r^*}, 
\text{as} \quad r\rightarrow \infty \label{sol}
\end{eqnarray}
and
\begin{eqnarray}
R^\infty_{lm\omega} &\sim& D^{\text{out}}e^{i k r^*}+\Delta^2 
D^{\text{in}} e^{-i
k r^*}, \text{as} \quad r\rightarrow r_+ \\
R^\infty_{lm\omega} &\sim& r^3 e^{-i \omega r^*}, \text{as} \quad 
r\rightarrow
\infty,
\end{eqnarray}
where $k=\omega-m a /(2 M r_+)$, $r_+=M+\sqrt{M^2-a^2}$ is the
radius of the black hole horizon, and $r^*$, the tortoise co-ordinate, is
defined as
\begin{equation}
r^*=r+{2 M r_+\over r_+-r_-}\ln{r-r_+\over2M}-{2M r_-\over r_+-r_-}
\ln{r-r_-\over 2M},
\end{equation}
where $r_-=M-\sqrt{M^2-a^2}$.

From Ref.\ \cite{Det}, the solution of the Teukolsky equation (solved
via a retarded Green's function) is
\begin{equation}
R_{lm\omega}(r)=R^\infty_{lm\omega}(r) Z^H(r)+
R^H_{lm\omega}(r) Z^\infty(r),
\end{equation}
where
\begin{equation}
Z^H(r)={1\over 2i\omega B^{\text{in}}_{lm\omega}}
\int^r_{r_+} {R^H_{lm\omega}(r) T_{lm\omega}(r)\over
\Delta^2} dr \label{z1}
\end{equation}
and
\begin{equation}
Z^\infty(r)={1\over 2 i \omega B^{\text{in}}_{lm\omega}}
\int^\infty_r {R^\infty_{lm\omega}(r) T_{lm\omega}(r)
\over \Delta^2} dr. \label{z2}
\end{equation}

For convenience, we will write $Z^H_{lm\omega}=Z^H(r\rightarrow\infty)$ and
$Z^\infty_{lm\omega}=Z^\infty(r\rightarrow r_+)$, and therefore our two
solutions can be written as 
\begin{equation}
R_{lm\omega}(r\rightarrow \infty) \sim Z^H_{lm\omega} r^3 e^{i \omega
r^*}
\end{equation}
and
\begin{equation}
R_{lm\omega}(r\rightarrow r_+) \sim Z^\infty_{lm\omega}
\Delta^2 e^{-i k r^*}.
\end{equation}

\section{Energy and angular momentum fluxes}

Towards infinity, the Weyl scalar can be related to the two fundamental
polarizations of gravitational waves by
\begin{equation}
\psi_4={1\over2}(\ddot{h}_+-i \ddot{h}_\times).
\end{equation}
From this and Eq.\ (\ref{weyl}) above, we can determine the averaged energy
and angular momentum fluxes at infinity, employing the Isaacson
stress-energy tensor to define the energy flux in the wave \cite{Is},
as 
\begin{equation}
\langle {dE_{\text{GW}}\over dt}\rangle = \dot{E}^\infty = \sum_{lmk} 
{|Z^H_{lmk}|^2\over 4\pi\omega_k^2} \label{ei}
\end{equation}
and
\begin{equation}
\langle {dL_{\text{GW}}\over dt}\rangle=\dot{L}_z^\infty=\sum_{lmk}
{m |Z^H_{lmk}|^2\over 4\pi \omega_k^3}, \label{li}
\end{equation}
where the amplitude coefficient is decomposed into a discrete set
of frequencies based on the particle's orbital motion,
\begin{equation}
Z^H_{lm\omega}=\sum_k Z^H_{lmk} \delta(\omega-\omega_k). \label{dec}
\end{equation}

Energy and angular momentum are also lost by radiation through the
horizon of the central black hole. Again, $\psi_4$ completely describes
the waves as $r^*\rightarrow -\infty$ and, with Teukolsky and Press 
\cite{TPIII}, we find 
\begin{equation}
\dot{E}^H=\sum_{lmk} \alpha^k_l {|Z^\infty_{lmk}|^2\over 4\pi\omega^2_k}
\label{eh}
\end{equation}
and
\begin{equation}
\dot{L}^H_z=\sum_{lmk} \alpha^k_l {m |Z^\infty_{lmk}|^2\over
4\pi\omega^3_k}, \label{lh}
\end{equation}
with an identical decompostion of $Z^\infty_{lm\omega}$ as with
$Z^H_{lm\omega}$, and where
\begin{equation}
\alpha^k_l={2^8 w_k^7 k (k^2+4 \epsilon^2)(k^2+16\epsilon^2)
(2 M r_+)^5 \over |C|^2} \label{alpha}
\end{equation}
and $\epsilon=\sqrt{M^2-a^2}/4 M r_+$ and
\begin{eqnarray}
|C|^2&=&[(\lambda+2)^2+4 a \omega m-4a^2\omega^2](\lambda^2+36 a
\omega m-36a^2\omega^2)\nonumber
\\ &+&(2\lambda+3)(96 a^2\omega^2-48a \omega m)
+144\omega^2(M^2-a^2).
\end{eqnarray}
The total rates of loss of energy and angular momentum by the
system are $\dot{E}^H+\dot{E}^\infty$ and
$\dot{L}^H_z+\dot{L}^\infty_z$.

\section{The Sasaki-Nakamura equation}
The preceding section makes it clear that our chief task is to
calculate the amplitudes $Z^{H,\infty}_{lmk}$, and it is apparent
from Eqs.\ (\ref{z1}) and (\ref{z2})
that this will entail solving the Teukolsky equation
to find the amplitude of the in-going waves at infinity, 
$B^{\text{in}}_{lm\omega}$
from Eq.\ (\ref{sol}). Numerically this presents a problem, however,
since the ingoing waves for this solution are completely swamped
by the outoing waves at large radii [compare amplitudes of $B^{\text{out}}
_{lm\omega} r^3$ and $B^{\text{in}}_{lm\omega}/r$ as $r\rightarrow \infty$
in Eq.\ (\ref{sol})]. In the Schwarzschild case this problem is
typically avoided by solving instead the Regge-Wheeler equation, and
transforming its solution to that of the Teukolsky equation via the
Chandrasekhar transformation \cite{Chandra4}. 
The virtue of this is that in the 
Regge-Wheeler formalism, with its short-range potential, 
the ingoing and outgoing waves near infinity
have the same order of magnitude.

In the Kerr case Sasaki and Nakamura have found an equation
with the same useful properties as the Regge-Wheeler equation in
Schwarzschild which, moreover, reduces to the latter equation when
$a\rightarrow 0$ \cite{SN}. The Sasaki-Nakamura equation is written as
follows
\begin{equation}
{d^2 X_{lm\omega}\over dr^2}-F(r){dX_{lm\omega}\over
dr}-U(r)X_{lm\omega}=0. \label{sn}
\end{equation}
The functions $F(r)$ and $U(r)$ are given in the appendix. The
equivalents to our two solutions to the Teukolsky equation are
\begin{eqnarray}
X^H_{lm\omega}&\sim& A^{\text{out}}_{lm\omega} e^{i\omega r^*}
+A^{\text{in}}_{lm\omega}e^{-i\omega r^*}, \text{as} \quad r\rightarrow \infty
\label{ain} \\
X^H_{lm\omega}&\sim& e^{-i k r^*}, \text{as} \quad r\rightarrow r_+
\label{aoth}
\end{eqnarray}
and
\begin{eqnarray}
X^\infty_{lm\omega}&\sim& e^{i \omega r^*}, \text{as} \quad r\rightarrow
\infty \\
X^\infty_{lm\omega}&\sim& D^{\text{out}} e^{i k r^*}+D^{\text{in}}e^{-i k
r^*}, \text{as} \quad r\rightarrow r_+.
\end{eqnarray}
The transformations between the quantities we require are
\begin{eqnarray}
R^H_{lm\omega}&=&{1\over \eta}[(\alpha+{\beta_{,r}\over\Delta})
\chi^H_{lm\omega}-{\beta\over\Delta}\chi^H_{lm\omega,r}], \label{t1}\\
R^\infty_{lm\omega}&=&-{c_0\over 4\omega^2 \eta}
[(\alpha+{\beta_{,r}\over\Delta})
\chi^\infty_{lm\omega}-{\beta\over\Delta}\chi^\infty_{lm\omega,r}]
\label{t2},
\end{eqnarray}
and
\begin{equation}
B^{\text{in}}_{lm\omega}=-{1\over 4\omega^2}A^{\text{in}}_{lm\omega}
\label{t3},
\end{equation}
where $\chi^{H,\infty}_{lm\omega}=X^{H,\infty}_{lm\omega} \Delta/\sqrt{r^2
+a^2}$, and $c_0$, $\alpha$, $\beta$ and $\eta$ are given in the
appendix.

\section{The source term}
The Teukolsky source term is given by \cite{B4}
\begin{equation}
T_{lm\omega}=4\int\rho^{-5}\bar{\rho}^{-1}(B_2^\prime+B_2^{\prime *})
e^{-i m \phi +i \omega t} \mbox{}_{-2} S^{a\omega}_{lm} d\Omega dt,
\end{equation}
where
\begin{eqnarray}
B_2^\prime&=&-{1\over2}\rho^8\bar{\rho}L_{-1}[\rho^{-4}L_0(\rho^{-2}
\bar{\rho}^{-1}T_{nn})] \nonumber \\ &-&
{1\over 2\sqrt{2}}\rho^8\bar{\rho}\Delta^2 L_{-1}
[\rho^{-4}\bar{\rho}^2J_+(\rho^{-2}\bar{\rho}^{-2}\Delta^{-1}T_{\bar{m}n})], \\
B_2^{\prime *}&=&-{1\over4}\rho^8\bar{\rho}J_+[\rho^{-4}J_+(\rho^{-2}
\bar{\rho}T_{\bar{m}\bar{m}}] \nonumber \\ &-&
{1\over 2\sqrt{2}}\rho^8\bar{\rho}\Delta^2 J_+
[\rho^{-4}\bar{\rho}^2\Delta^{-1}L_{-1}(\rho^{-2}\bar{\rho}^{-2}T_{\bar{m}n})],
\end{eqnarray}
and $\rho=(r-i a \cos\theta)^{-1}$, with $\bar{\rho}$ its complex
conjugate. The operators $L_s$ and $J_+$ are defined as
\begin{equation}
L_s=\partial_\theta+{m \over \sin\theta}-a\omega\sin\theta+s\cot\theta
\end{equation}
and
\begin{equation}
J_+=\partial_r+i{K\over \Delta}.
\end{equation}

The tetrad components of the particle's energy momentum tensor can
be written
\begin{eqnarray}
T_{nn}&=& {C_{nn}\over \sin\theta}\delta(r-r(t))\delta(\theta-\pi/2)
\delta(\phi-\phi(t)), \\
T_{\bar{m}n}&=& {C_{\bar{m}n}\over \sin\theta}\delta(r-r(t))\delta(\theta-\pi/2)
\delta(\phi-\phi(t)), \\
T_{\bar{m}\bar{m}}&=& {C_{\bar{m}\bar{m}}
\over \sin\theta}\delta(r-r(t))\delta(\theta-\pi/2)
\delta(\phi-\phi(t)), 
\end{eqnarray}
where
\begin{eqnarray}
C_{nn}&=&C^{(0)}_{nn}+C^{(1)}_{nn} {dr\over dt}+ 
C^{(2)}_{nn} ({dr\over dt})^2 \nonumber \\
&=&{\mu\over 4\Sigma^3 \dot{t}}[E(r^2+a^2)-a L_z]^2+{\mu \over
2 \Sigma^2}[E(r^2+a^2)-a L_z] {dr\over dt} \nonumber \\
&+&{\mu\dot{t}\over 4
\Sigma} ({dr\over dt})^2\\
C_{\bar{m}n}&=&C^{(0)}_{\bar{m}n}+C^{(1)}_{\bar{m}n} {dr\over dt}
\nonumber \\ &=&
{\mu \rho \over 2\sqrt{2}\Sigma^2 \dot{t}}
[E(r^2+a^2)-a L_z][i\sin\theta(a E-{L_z\over \sin^2\theta})] 
\nonumber \\ &-&{\mu \rho \over 2\sqrt{2}\Sigma}[i\sin\theta(a E-{L_z\over
\sin^2\theta}] {dr\over dt}\\
C_{\bar{m}\bar{m}}&=&{\mu \rho^2 \over 2\Sigma \dot{t}}
[i\sin\theta(a E-{L_z\over \sin^2\theta})]^2
\end{eqnarray}
and $\dot{t}=dt/d\tau$.

Integrating by parts, and making use of the adjoint operator 
$L_s^\dagger=\partial_\theta-m/\sin\theta+a\omega\sin\theta
+s \cot\theta=\partial_\theta +f(\theta)$, 
which bears the following useful relation to the
operator $L_s$ defined above:
\begin{equation}
\int^\pi_0 h(\theta) L_s[g(\theta)]\sin\theta d\theta=
-\int^\pi_0g(\theta)L^\dagger_{1-s}[h(\theta)]\sin\theta d\theta,
\end{equation}
with $g(\theta)$ and $h(\theta)$ arbitrary functions \cite{tag},
we find that
\begin{eqnarray}
T_{lm\omega}&=& \int^\infty_{-\infty} \int^{2\pi}_0
\Delta^2[\{(A_{nn0}+A_{\bar{m}n0}
+A_{\bar{m}\bar{m}0})\delta(r-r(t)\} \nonumber \\
&+&\{(A_{\bar{m}n1}+A_{\bar{m}\bar{m}1})
\delta(r-r(t))\}_{,r} \nonumber \\ &+&
\{A_{\bar{m}\bar{m}2} \delta(r-r(t))\}_{,rr}]
\delta(\phi-\phi(t)) e^{i\omega t-i m \phi} d\phi dt .
\end{eqnarray}
The $A$'s are all functions of $r$ only, and in each case 
$A=A^{(0)}+A^{(1)}(dr/dt)+A^{(2)}(dr/dt)^2$,
where
(writing $\mbox{}_{-2} S^{a\omega}_{lm}$ simply as $S$ hereafter for simplicity)
\begin{eqnarray}
A^{(i)}_{nn0}&=&-{2\over \Delta^2}C^{(i)}_{nn} r^3 (r S_{,\theta\theta}-2 i a
S_{,\theta}+2 r f(\pi/2)S_{,\theta} \nonumber \\ &-& 2i a f(\pi/2) S
+r S (f(\pi/2)^2-2), \\
A^{(i)}_{\bar{m}n0}&=&{2\sqrt{2}\over\Delta}C^{(i)}_{\bar{m}n}r^3(S_{,\theta}+
f(\pi/2)S)(i {K\over\Delta}+{2\over r}), \\
A^{(0)}_{\bar{m}\bar{m}0}&=&-r^2 C_{\bar{m}\bar{m}} S(-i ({K\over\Delta})_{,r}
-({K\over\Delta})^2+{2i\over r}{K\over\Delta}), \\
A^{(i)}_{\bar{m}n1}&=&{2\sqrt{2}\over\Delta}r^3 C^{(i)}_{\bar{m}n}(S_{,\theta}
+f(\pi/2)S), \\
A^{(0)}_{\bar{m}\bar{m}1}&=&-2 r^2 C_{\bar{m}\bar{m}} S(i{K\over\Delta}
+{1\over r}), \\
A^{(0)}_{\bar{m}\bar{m}2}&=&-r^2 C_{\bar{m}\bar{m}} S, \\
A^{(2)}_{\bar{m}n0}&=&A^{(2)}_{\bar{m}n1}=A^{(1)}_{\bar{m}\bar{m}0}=
A^{(2)}_{\bar{m}\bar{m}0}=A^{(1)}_{\bar{m}\bar{m}1}=A^{(2)}_{\bar{m}
\bar{m}1}=A^{(1)}_{\bar{m}\bar{m}2}=A^{(2)}_{\bar{m}\bar{m}2}=0 .
\end{eqnarray}
In every case the spheroidal harmonic function ($S$) 
and its derivatives are evaluated
at $\theta=\pi/2$.

It is now easy to show, from Eqs.\ (\ref{z1}) and (\ref{z2}) 
and using integration
by parts (keeping in mind that we are interested only in closed
orbits, for which $r_+<r<\infty$ always holds strictly), that
\begin{equation}
Z^{H,\infty}_{lm\omega}={1\over 2i\omega B^{\text{in}}_{lm\omega}}
\int_{r_+}^\infty \int^\infty_{-\infty} \int^{2\pi}_0
I^{H,\infty}_{lm\omega}(r) \delta(r-r(t))\delta(\phi-\phi(t)) d\phi dt dr,
\label{key}
\end{equation}
for which $I^{H,\infty}_{lm\omega}(r)=I^{(0)}_{lm\omega}(r)+
I^{(1)}_{lm\omega}(r) (dr/dt)+I^{(2)}_{lm\omega}(r) (dr/dt)^2$, where
\begin{equation}
I^{(i)}_{lm\omega}(r)=
R^{H,\infty}_{lm\omega}(A^{(i)}_{nn0}+A^{(i)}_{\bar{m}n0}+
A^{(i)}_{\bar{m}\bar{m}0})
-{d R^{H,\infty}_{lm\omega}\over dr}(A^{(i)}_{\bar{m}n1}+
A^{(i)}_{\bar{m}\bar{m}1})
+{d^2 R^{H,\infty}_{lm\omega}\over dr^2} A^{(i)}_{\bar{m}\bar{m}2}
\label{II}.
\end{equation}

It is necessary to expand $Z^{H,\infty}_{lm\omega}$ in terms of
the eccentricity $e$, keeping in mind that we wish, as shown in
section 2 above, to find $\dot{E}^{H,\infty}$ and
$\dot{L}_z^{H,\infty}$ to second order in $e$, and that each of these
is proportional to $|Z^{H,\infty}_{lm\omega}|^2$. However, it
transpires that only terms up to order $e$ in the integrand of Eq.\
(\ref{key})
contribute to order $e^2$ in $\dot{e}$, the rate of change of
eccentricity derived from $\dot{E}^{H,\infty}$ and $\dot{L}_z
^{H,\infty}$. The reasons for this
emerge as we proceed to expand $I^{H,\infty}_{lm\omega}(r)$,
$\delta(r-r(t))$ and $\delta(\phi-\phi(t))$ in powers of $e$.

Employing the expansions of $r(t)$ and $\phi(t)$ derived above
[Eqs.\ (\ref{R}) and (\ref{Phi})], we can write the product of
delta functions in Eq.\ (\ref{key}) as a product of two Taylor
expansions in the small parameter $e$, about the points
$r-r_0$ and $\phi-\Omega_\phi t$.
\begin{eqnarray}
\delta(r-r(t))\delta(\phi-\phi(t))&=&\delta(r-r_0)\delta(\phi-\Omega_\phi
t)-e r_0 \cos \Omega_r t \delta^\prime(r-r_0)\delta(\phi-\Omega_\phi t)
\nonumber \\
&-&e p(r_0) \sin \Omega_r t \delta^\prime(\phi-\Omega_\phi t)\delta(r-r_0)
+O(e^2),
\end{eqnarray}
where the prime denotes differentiation with respect to
the function's argument. We can 
integrate by parts in Eq.\ (\ref{key}) to integrate terms containing
derivatives of delta functions, and this will simply mean that 
$\delta^\prime (\phi-\Omega_\phi t)$ will be replaced by 
$i m \delta(\phi-\Omega_\phi t)$, since $e^{-i m \phi}$ is the only
other part of the integrand which depends on $\phi$. Completing the
$\phi$ integration thus leaves us with the overall factor
$e^{i(\omega-m \Omega_\phi)t}$, and some terms depending on $\cos
\Omega_r t$, $\sin \Omega_r t$ and, in the $O(e^2)$ part, on
$\cos 2 \Omega_r t$ and $\sin 2 \Omega_r t$. Following the time
integration, then, we will have a series of delta functions of
the type $\delta(\omega-m \Omega_\phi)$ [at all orders, except
$O(e)$], $\delta(\omega-m \Omega_\phi \pm \Omega_r)$ (at all orders
from $O(e)$ up) and, in general, $\delta(\omega-m \Omega_\phi \pm
k \Omega_r)$ at $O(e^k)$ and above. 

These delta functions, after
integration over $\omega$ to derive $\psi_4$ [Eq.\ (\ref{weyl})],
produce terms representing energy and angular momentum emitted
at the fundamental (circular motion) frequency $w_m=m \Omega_\phi$,
and at a series of discrete sidebands, $w_\pm=m \Omega_\phi \pm
\Omega_r$ and $w_{\pm k}=m \Omega_\phi \pm k \Omega_r$. The 
occurrence of these
delta functions justifies the decomposition of $Z^{H,\infty}_
{lm\omega}$ referred to earlier [Eq.\ (\ref{dec}) above]. 

It is, of course, $|Z^{H,\infty}_{lm\omega}|^2$ which is integrated
in Eq.\ (\ref{weyl}). Therefore, up to order $e^2$, only those
$O(e^2)$ terms in $Z^{H,\infty}_{lm\omega}$ which cross multiply
with $O(e^0)$ terms will contribute. Since the frequency must be
single valued for any given term, only the circular harmonic ($w_m$)
term in $O(e^2)$ survives the Fourier transform which produces the
Weyl scalar, all other terms being annihilated. The $O(e)$ terms
in $Z$ have no circular harmonic term, as mentioned before, so
these terms only contribute to loss of energy and engular momentum
at $O(e^2)$. 

As seen from Eq.\ (\ref{edotf}) above, it is the difference $\dot{E}_2-
\Omega \dot{L}_2$ on which $\dot{e}$ actually depends at leading
order. Eqs.\ (\ref{ei}),(\ref{li}),(\ref{eh}) and (\ref{lh}) show that 
\begin{equation}
\dot{E}_n-\Omega \dot{L}_n \propto
1-{m \Omega \over \omega_k}, \quad \text{at order} \quad e^n
\end{equation}
which is zero to leading order if $\omega_k=\omega_m=m\Omega_\phi$.
This means not only that 
the $O(e^2)$ terms in 
$Z^{H,\infty}_{lm\omega}$ do not contribute at all to $\dot{e}$
below $O(e^3)$, but also that $\dot{E}_0-\Omega \dot{L}_0$ is also
zero to leading order, as noted above [Eq.\ (\ref{cc})]. In fact,
since the eccentric correction to the azimuthal frequency $\Omega_\phi$
is itself of $O(e^2)$, the circular losses of energy and angular
momentum contribute to $\dot{e}$ at $O(e^2)$ to leading order,
like the first order terms in $Z$.
Therefore there is no loss of $E$ and $L_z$
at $O(e)$, and so $\dot{E}_1=0$ and $\dot{L}_1=0$ 
as claimed in section 2.

This proves that a precisely circular equatorial orbit in
Kerr will always remain circular under radiation reaction (as long
as the adiabatic approximation still holds). Furthermore it means
that to find the leading order correction to this condition for slightly
eccentric orbits, and thus establish the stability of circularity,
we need only examine the $O(e)$ terms in Eq.\ (\ref{weyl}), and
can drop all $O(e^2)$ corrections to the motion, except for the
$\Delta \Omega$ part of $\Omega_\phi$. This also means, of course, that
only 
contributions to the loss of energy and angular momentum from the
first pair of sidebands ($\omega=\omega_\pm$) need be included
with the circular harmonic ($\omega_m$) in calculating $\dot{e}$
to leading order.

\section{Calculation of rate of change of eccentricity}

As a final step before integration of Eq.\ (\ref{key}), 
the function $I^{H,\infty}_{lm\omega}(r)$
must be expanded up to first order in $e$. It contains terms which
depend on $dr/dt$ which, by Eq.\ (\ref{R}) above, is $O(e)$ at
leading order, $dr/dt=-e r_0 \Omega_r \sin \Omega_r t+O(e^2)$.
Therefore we will write
\begin{equation}
I^{H,\infty}_{lm\omega}(r)=I^{(0)}_{lm\omega}(r)-e I^{(1)}_{lm\omega}
(r) r_0 \Omega_r \sin\Omega_r t+O(e^2).
\end{equation}
Thus, doing a final integration by
parts in the integral over $r$ in Eq.\ (\ref{key}), we find
\begin{equation}
Z^{H,\infty}_{lm\omega}=-{\pi\over i \omega B^{\text{in}}_{lm\omega}}
[I^{(0)}_{lm\omega}(r_0)\delta(\omega-m\Omega_\phi)-
e B^+_{lm} \delta(\omega-\omega_+)- e B^-_{lm}
\delta(\omega-\omega_-)+O(e^2)], \label{zz}
\end{equation}
where
\begin{equation}
B^\pm_{lm}={1\over2}(r_0 {dI^{(0)}_{lm\omega}\over dr}\bigg\vert_{r=r_0}
\pm m p(r_0) I^{(0)}_{lm\omega}(r_0) \mp I^{(1)}_{lm\omega}(r_0)
r_0 \Omega_r). \label{bpl}
\end{equation}

The argument of the preceding section shows that,
in order to calculate the quantity $\dot{E}_2-\Omega\dot{L}_2+
\Delta \Omega \dot{E}_0$, we need only evaluate the co-efficients
$B^\pm_{lm}$ in $Z_{lm}$. Therefore, returning to Eq.\ (\ref{edotf}),
we have 
\begin{equation}
\mu\dot{e}/e=- j(v)[\Gamma-h(v) \dot{E}_0] \label{final}
\end{equation}
where
\begin{eqnarray}
\Gamma&=&\dot{E}_2-\Omega\dot{L}_2+\Delta\Omega \dot{E}_0 \\
&=& {\Omega_r \over 4 \pi}  \sum_{lm} \Bigl({|B^{H+}_{lm}|^2\over \omega_+^3}-
{|B^{H-}_{lm}|^2\over\omega_-^3}\Bigr)
\nonumber \\ &+&{\Omega_r\over 4\pi}\sum_{lm}
\Bigl({|B^{\infty +}_{lm}|^2\over \omega_+^3} \alpha^+_l
-{|B^{\infty -}_{lm}|^2\over \omega_-^3} \alpha^-_l\Bigr) \label{hun}
\end{eqnarray}
and
\begin{equation}
h(v)=\Delta\Omega-g(v)={{\cal H}(v)
\over2(1+q v^3)(1-2v^2+q^2v^4)^2(1-6v^2+8q v^3-3q^2v^4)},
\label{hv}
\end{equation}
with
\begin{eqnarray}
{\cal H}(v)=
1&-&12v^2+66v^4-108v^6+q v^3+8q^2v^4-72q v^5
-20q^2v^6 \nonumber \\ 
&+&204q v^7+38q^3v^7-42q^2v^8-9q^4v^8-144q^3v^9+116q^4v^{10}
\nonumber \\ &-& 27q^5v^{11}. \label{calH}
\end{eqnarray}

As an aside, we take the opportunity to write the eccentricity in
terms of quantities which can be deduced from the signal observed
in a detector such as LISA. The complex wave amplitude at earth
$h(t)=h_+-i h_\times$ can be written as 
\begin{equation}
h(t)=-{r^3 \over (r-i a \cos\theta)^4} \int^\infty_{-\infty} 
\sum_{lm} {1 \over \omega^2} Z^H_{l m \omega} 
\mbox{}_{-2} S^{a\omega}_{lm}(\theta) e^{i m \phi} e^{-i
\omega(t-r^*)} d\omega , 
\end{equation}  
where $r$ is the distance from the source to Earth and $t-r^*$
is retarded time. 
A glance at Eq.\ (\ref{zz}) tells us we can define, based on
this equation, amplitudes for the main sideband with frequency
$\omega_m$ (call this amplitude $h_m$) and for the various
sidebands (let $h_1$ be the amplitude for the sideband of frequency
$\omega_+$). To leading order $h_m$ will not depend on the eccentricity,
whereas $h_1$, the amplitude of the first sideband, will be linear
in $e$. It is therefore easy to show that the eccentricity will be
proportional to the ratio of the amplitudes of the first and the
main sidebands (i.e. $h_1/h_m$). In fact,
\begin{equation} 
e=\bigg\vert {h_1 \over h_m} \bigg\vert {\sum {1\over\omega^3_m}
{1\over B^{\text{in}}_{lm\omega_m}} I^{(0)}_{lm\omega_m}(r_0) 
\mbox{}_{-2} S^{a\omega_m}_{lm}(\theta) e^{i m \phi} e^{-i
\omega_m (t-r^*)} \over
\sum{1\over\omega^3_+} 
{1\over B^{\text{in}}_{lm\omega_+}} B^+_{lm\omega_+} (r_0)     
\mbox{}_{-2} S^{a\omega_+}_{lm}(\theta) e^{i m \phi} e^{-i
\omega_+ (t-r^*)}} \label{edef2}.
\end{equation} 
In order to measure $e$ as it evolves with the signal, the signal
will have to be strong enough to permit not only measuring the size
of the first sideband, but also some parameter extraction, 
so that $a$ and $M$ can be estimated. 

In summary, Eq.\ (\ref{final}) is the equation which allows us to
compute the change in eccentricity for an inspiralling orbit, and
Eq.\ (\ref{rdot}) defines the rate of inspiral. Eq.\ (\ref{hun}),
Eq.\ (\ref{bpl}) and Eq.\ (\ref{II}), for $\Gamma$, and
Eqs.\ (\ref{ei}) and (\ref{eh}) with the $O(e^0)$ part of Eq.\ (\ref{zz})
for $\dot{E}_0$, allow us to express $\dot{e}$ in terms of
the solution of the radial Tuekolsky equation $R^{H,\infty}_{lm\omega}$,
and its derivatives, as well as the incoming wave amplitude
$B^{in}_{lm\omega}$. These quantities are in turn derived numerically
by solving the Sasaki-Nakamura equation as described below in
section 9, and employing the transformations given in Eqs.\ (\ref{t1}),
(\ref{t2}) and (\ref{t3}). The important functions $j(v)$,
$h(v)$ and $\Delta\Omega$ in Eq.\ (\ref{final}) are all derived
in solving the equations of geodesic motion for the orbiting body
in section 2.

\section{Adiabatic condition}
The whole preceding argument depends on an adiabatic condition on
the motion which says that the inspiral timescale $r_0/|\dot{r}_0|$
is much greater than the orbital period of the motion $2\pi/\Omega_r$.
The necessity for this condition is most noticeable in the approximation
which describes the evolution of the particle's motion under back
reaction as passing through a series of geodesic orbits, each defined
as if no back reaction were taking place during that orbit. Once the
inspiral proceeds on a timescale which is about as short as the time
to complete an orbit, this approximation loses all validity. Using
Eq.\ (\ref{rdot}), we find that the adiabatic condition can be written,
\begin{equation}
{\mu\over M}\ll{v^5\over 2\pi}{(1-6v^2+8q v^3-3q^2v^4)^{3/2}\over
(1-3v^2+2q v^3)^{3/2} (1+q v^3)} {1\over (M/\mu)^2 \dot{E}_0}.
\label{Ad}
\end{equation}
Just as the inspiral timescale must be greater than an orbital
period, so too must the circularization timescale $e/\dot{e}$.
However, this quantity is almost invariably less than the
inspiral timescale, so Eq.\ (\ref{Ad}) is the key condition.
For very large radii, in the Newtonian limit, $(M/\mu)^2 \dot{E}_0
\simeq 32 v^{10}/5$ (for a discussion of this limit see Ref.\
\cite{III})
and the condition is simply $\mu/M\ll (5/128\pi)v^{-5}$, which
is very much less restrictive than the linear perturbation condition
$\mu/M\ll 1$, upon which the Teukolsky formalism rests. Approaching
the ISCO however, where the numerical results tell us that 
$\dot{E}_0$ remains finite and of the same order as its
Newtonian value, we see that the adiabatic limit on $\mu/M$ is
proportional to 
$(1-6v^2+
8q v^3-3q^2v^4)^{3/2}$, which becomes vanishingly small as the ISCO
nears. Therefore, near the ISCO the adiabatic condition supercedes
the linear perturbation condition, as the leading constraint
on $\mu/M$. Only by imagining a test particle
which has vanishingly small mass can we apply the results of
our calculation all the way to the ISCO, but no doubt there
exist real physical systems, with $\mu/M\leq 10^{-6}$ for
instance, which are
correctly described for almost all of the inspiral by this approximation
(recalling that our calculations presume that the particle is a point
mass as a further simplification). This issue will be discussed more
quantitatively in Ref.\ \cite{FOT}.

\section{Results}

With the results of section 7, it only remains to calculate
$R^{H,\infty}_{lm\omega}$, $B^{\text{in}}_{lm\omega}$ [Eqs.\
(\ref{sola})
and (\ref{sol})] and
$\mbox{}_{-2} S^{a\omega}_{lm}(\pi/2)$ [Eq.\ (\ref{weyl})] numerically
to find $\dot{e}/e$. To find the solutions to the radial equation
[Eq.\ (\ref{Tr})]
one actually solves the Sasaki-Nakamura equation [Eq.\ (\ref{sn})]
for $X^{H,\infty}_{lm\omega}$ and $A^{\text{in}}_{lm\omega}$
[Eqs.\ (\ref{ain}) and (\ref{aoth})]. These solutions are very
smooth, apart from a singularity at the horizon $r_+$, and so
Bulirsch-Stoer integration works very well in integrating them.
The singularity is avoided by starting the integration from
a point just outside the horizon (typically at $r_+ + 10^{-8}$).
The solutions are insensitive to variations by several orders
of magnitude of this small increment. 
Richardson polynomial extrapolation is used to evaluate
$A^{\text{in}}_{lm\omega}$ as $r\rightarrow\infty$,
since it can be expressed as the first term in a polynomial
in $1/\omega r$ defining the amplitude of the ingoing wave
at large $r$ in Eq.\ (\ref{ain}) \cite{gen}. This amplitude
is evaluated for several endpoints of integration, doubling
the endpoint radius at each trial, allowing the extrapolator
to evaluate the limit of the amplitude as $r\rightarrow \infty$,
which is $A^{\text{in}}_{lm\omega}$. 

The Spheroidal harmonic functions are calculated by 
expressing them as a linear combination of spherical harmonics
of equal $m$, summed over all available values in $l$ (truncating
the series after 30 terms in practice).
Substituting this series into the second-order ODE defining the
spheroidal harmonics gives us a 5-term recurrence relation for
the co-efficients of the expansion. 
This procedure, for the scalar case only, is found in \cite{MF4}.
The recurrence relation for the expansion co-efficients can be solved using
matrix eigenvalue routines which, like the Bulirsch-Stoer
integrator
and the polynomial extrapolator, 
are found in Ref.\ \cite{NumRec4}. The derivative of each spheroidal
harmonic is also expressible as a combination of spherical
harmonics of different spin-weight values by use of the edth operator
\cite{Gold4}.
Useful checks for the numerical results are found in the Schwarzschild
limit, in \cite{III} and in the circular limit, in 
\cite{Shib}. Analytically
the results of sections 2 and 7 
reduce to those of \cite{III} in the Schwarzschild limit
and those of section 2 to the results 
of \cite{tag} in the post-Newtonian limit. 

The accuracy of the numerical results is limited by several 
factors. The relative accuracies of the Bulirsch-Stoer integrator and
the Richardson extrapolator can be increased easily, at some
loss in computing speed. For these calculations they were
set to $10^{-6}$ and $10^{-5}$ 
respectively. The solution of the eigenvalue problem
has very good accuracy, but the approximation of the spheroidal
harmonics as a combination of spherical harmonics begins to lose
accuracy seriously when $a \omega$ becomes much larger than order
unity. However, this only occurs for very high ($m>20$) harmonics
of the motion for small radii, and these contributions are not
required at the accuracy used here. The chief limit on accuracy
is, in fact, the number of harmonics in $l$ and $m$ which are
calculated. Invariably, for small eccentricity orbits, the leading
order contribution is for $l=2$, $m=2$, and the significance of
the contribution decreases sharply (but less so for small radii)
with increasing $l$ and $m$. A simple estimate, used in Ref.\
\cite{III}, enables one to reliably estimate the inaccuracy involved
in truncating the calculation at $l=l_{\text{max}}$. It tells
us that, for a relative error (in estimates of the loss of
energy and angular momentum) no greater than $\eta$, with
a mean orbital radius $r_0$, then $l_{\text{max}}\geq 
\log \eta/ \log(M/r_0)+3$. Taking all of these factors into
account, we can generally estimate the accuracy of the numerical results 
at $10^{-4}$, and certainly the relative errors should be no
greater than $10^{-3}$ in the worst case. 

A useful parameter with which to investigate the orbital evolution
is $c$, which represents a ratio of the inspiral timescale to
the circularization timescale, or
\begin{equation}
c={r_0\over e} {de/dt \over dr_0/dt} \label{cee}.
\end{equation}
Again, $c$ is positive when radiation reaction circularizes
the orbit, and negative when it drives the orbit more eccentric.
In order to see analytically the behaviour of $c$ as the ISCO approaches,
recall Eq.\ (\ref{final}) and write
\begin{equation}
c=-{r_0\over \mu \dot{r}_0} j(v)[\Gamma-h(v)\dot{E}_0].
\end{equation}
As $r_0\rightarrow r_{\text{ISCO}}$, the radius of the innermost
stable circular orbit, the function $h(v)$ [Eq.\ (\ref{hv})] diverges,
since $r_{\text{ISCO}}^2-6 M r_{\text{ISCO}}+8 a \sqrt{M r_{\text{ISCO}}}
-3 a^2=0$.
Since the
numerical results show that $\Gamma$ and $\dot{E}_0$ remain finite 
in all cases,
it is apparent that $\Gamma$ (which is otherwise dominant), contributes
negligibly near $r_{\text{ISCO}}$. Therefore, making use of
the expression for $\dot{r}_0$ from Eq.\ (\ref{rdot}), we find
for $r_0$ near $r_{\text{ISCO}}$,
\begin{equation}
c\sim-{{\cal H}\over 4
(1-2v^2+q v^3)(1-3v^2+2qv^3)(1-6v^2+8q v^3-3q^2 v^4)}. \label{ff}
\end{equation}
Again, $1-6v^2+8q v^3-3q^2v^4\rightarrow 0$ as $r\rightarrow r_{\text
{ISCO}}$, so $c$ diverges at the ISCO. However, its sign as this point
approaches depends on the function ${\cal H}$ [Eq.\ (\ref{calH})], since
the expressions in the denominator are all positive for
$r>r_{\text{ISCO}}$.
It is obvious that for large $r$,
${\cal H}$ is always positive, but for small values of $r$, which can
be achieved by
prograde orbits around rapidly
spinning black holes ($a>.95M$), ${\cal H}$ can become negative. However,
it always becomes positive again before the ISCO, so that 
$c\rightarrow -\infty$ at the ISCO, in all cases except one.

The exceptional case is 
the extreme one of $a\rightarrow M$. 
At this unique point, ${\cal H}$,
and all expressions in the denominator of Eq.\ (\ref{ff}) go to zero.
Setting $q=1$ in Eq.\ (\ref{ff}), and canceling factors of $(v-1)$
from both numerator and denominator, one finds that 
\begin{equation}
\lim_{q=1,v\rightarrow 1} c=3/2,
\end{equation}
which is both positive and finite, in contrast to the usual behaviour
as the ISCO approaches. 

As Fig. 1 shows, the curves describing the critical radius
and the ISCO do approach each other in terms of the Boyer-Lindquist
radial coordinate as 
$a\rightarrow M$, as our analysis of $c$ might suggest. Therefore
it is interesting to investigate the consequences of this for massive
particles inspiraling around near extreme Kerr black holes. A useful
measure here is the number of orbits left in the inspiral once the
particle reaches the critical radius, that is, the number of orbits
it will take the particle to reach the ISCO. Defining $t_c$ as the
inspiral time between $r_{\text{crit}}$ and $r_{\text{ISCO}}$, and referring
to Eq.\ (\ref{rdot}) for the rate of inspiral, we have
\begin{equation}
t_c=-\int^{r_{\text{ISCO}}}_{r_{\text{crit}}} {v^2 (1-6v^2+8q v^3-3q^2v^4)\over
2 \dot{E}_0(1-3v^2+2q v^3)} {\mu d r_0\over r_0}.
\end{equation}
To a rough approximation, we can take $\dot{E}_0$ as constant in
this region, and therefore
\begin{equation}
t_c\approx {\mu\over \dot{E}_0}
\bigg\vert\sqrt{1-3v^2+2q v^3}\biggl(-{1\over2}+
{v^2-1\over2(1-3v^2+2q v^3)}\biggr)\bigg\vert^{v_{\text{ISCO}}}_
{v_{\text{crit}}}.
\end{equation}
Approximately, the number of orbits left in this time will be
\begin{equation}
N_c\approx {t_c\over T} \approx {t_c \Omega\over 2 \pi}\approx
{\mu\over M} {v_{\text{crit}}^3\over 2 \pi \dot{E}_0}{1\over1+q 
v^3_{\text{crit}}}
\bigg\vert\sqrt{1-3v^2+2q v^3}\biggl(-{1\over2}+
{v^2-1\over2(1-3v^2+2q v^3)}\biggr)\bigg\vert^{v_{\text{ISCO}}}_
{v_{\text{crit}}}.
\end{equation}
Note that $\dot{E}_0\propto (\mu/ M)^2$, so that $N_c$ is inversely
proportional to $\mu/M$. In the test particle limit
$\mu/M\rightarrow 0$, $N_c\rightarrow
\infty$.

For $a=-.9 M$, we find that $N_c\approx .035 M/\mu$, while for
$a=.99 M$, $N_c\approx .0025 M/\mu$. Note that the rate of energy loss
is similar in these two cases (retrograde orbits radiate more
energy for an orbit of given radius than do prograde orbits), 
but the distance between $r_{\text{crit}}$ and
$r_{\text{ISCO}}$ is much smaller in the latter case. The condition
of Eq.\ (\ref{Ad}) at the critical radius for $a=.99 M$ is
$\mu/M \ll .01$, so these estimates are still applicable to systems
with extreme mass ratios, such as compact solar-mass-size objects
spiralling into rapidly rotating 
supermassive black holes. For such a system, a
prograde orbit spends an order of magnitude or more fewer orbits in the
eccentricity increasing phase than does a retrograde orbit. Furthermore,
the orbital periods for these two cases (a prograde orbit with $r_0
\sim 1.5 M$, and a retrograde orbit with $r_0\sim 9.5M$) are also
very different, with the period of the retrograde orbit an order of
magnitude longer. The retrograde orbit therefore spends a factor of
hundreds more time gaining eccentricity than the prograde orbit. Conversely,
the prograde orbit spend much longer in the eccentricity decreasing
phase. 

Fig. 1 illustrates the positions of the horizon, ISCO and the
critical radius
for prograde and retrograde orbits around black holes of all spins.
Fig. 2 illustrates the behaviour of $c$ for Schwarzschild orbits 
($a=0$) and for
prograde and retrograde orbits around a Kerr black hole with $a=.9M$.
The dramatic plunge in $c$ towards negative values as the ISCO
approaches is seen in all three cases. 

In order to calculate how the eccentricity changes as the orbit evolves,
one can integrate Eq.\ (\ref{cee}) and define a new parameter $\gamma$,
such that
\begin{equation}
\gamma\equiv\ln \bigl({e_f \over e_i}\bigr)=
\int^{r_f}_{r_i} {c\over r_0} dr_0,   
\end{equation}
where $e_i$ is the initial eccentricity at radius $r_i$, and $e_f$ is the
eccentricity at a smaller radius $r_f$. Employing the numerical results
for $c$, along with the analytic approximation close to the ISCO given
in Eq.\ (\ref{ff}), we can numerically integrate this equation to
derive $\gamma$. Fig. 3 shows
$\gamma$ for the three cases of Fig. 2, illustrating how the eccentricity
changes as the orbit inspirals. One can see that, in the case of a black
hole with spin parameter $a=0.9 M$, a retrograde orbit ($q=-0.9$) will
have an order-of-magnitude greater eccentricity (relative to the eccentricity
it had at $r_0=100 M$) when it reaches the
critical radius (the turning point on the curve) then will a prograde
orbit when it reaches its critical radius, much further in.  
The amount the eccentricity increases by after the critical radius
is passed depends crucially on the details of the physical size and mass
of the orbiting particle, which it is beyond the scope of this paper
to analyse. For a test particle with vanishing $\mu/M$ the eccentricity
increases arbitrarily, but in a physical case this process will be cut
off by the onset of dynamical instability at some point. 

\section{Conclusions}

The results of this paper broadly confirm the experience of the
non-rotating case, in that radiation reaction tends to reduce orbital
eccentricity until near the the ISCO, when the onset of dynamical
instability is prefigured by a period of decircularization of the
inspiralling orbit. It seems reasonable to suppose that this effect
is induced by alterations in the shape of the radial potential $R$
as the ISCO approaches, since at the ISCO, the minimum which defines
the particle's circular orbit disappears. Beyond this point the particle
can only plunge towards the central body and is not longer in a dynamically
stable orbit. One can imagine that as this point approaches, the potential
well in which the orbit sits becomes shallower and broader (as it turns
into a saddle point), so that the orbital eccentricity increases despite
the circularizing force which drives the orbit towards the potential
minimum. The tendancy 
of prograde orbits around rapidly rotating black holes to begin
increasing in eccentricity only very shortly before the plunge
into the black hole (at $r_{\text{ISCO}}$)
suggests that massive bodies in such orbits will have smaller 
eccentricities at the end of their inspiral than 
with bodies in retrograde orbits, or the non-rotating case. In
the case of prograde orbits around an extreme Kerr black hole, 
the fact that $c$ is positive arbitrarily close to $r=M$, suggests
that the critical radius has descended in the ``throat'' of the black
hole along with the ISCO, a region where the Boyer-Lindquist co-ordinates
become degenerate. Since our notion of circularity is so dependant on
this co-ordinate system, it is unclear whether we can attach any meaning
to the critical radius for nearly circular orbits in this extreme
context. Nevertheless, from a practical point of view this critical radius
continues to be distinguishable, in terms of the B-L radius, 
from the radius of the ISCO as we approach
arbitrarily close to the case of extremal rotation, albeit that it
approaches the latter more and more closely as the rotation increases (for
prograde orbits). It is worth empasizing that our definition of the
eccentricity, although closely tied to a particular co-ordinate system,
is nevertheless an important {\it observable} element of the gravitational
wave signal emitted by the system, as seen above in Eq.\ (\ref{edef2}).

Another effect of the back reaction force on the orbit is one which
tends to alter the inclination angle, which measures the maximum departure
of the orbit from the equatorial plane. Ryan \cite{Fin2} has shown 
that nearly equatorial prograde orbits tend to increase their
inclination
angle under radiation reaction, thus moving away from
being equatorial, although the effect is not
very pronounced. Retrograde orbits, on the other hand, 
tend to decrease their inclination angle
(since the spin-orbit interaction is attractive for
retrograde orbits). Therefore, by the late stages of inspiral, one
might not expect prograde orbits to have remained very close to
the equatorial plane. This illustrates the need for a more general
calculation of orbital evolution in the Kerr geometry, which deals
with the issue of the Carter constant. 

\section*{Acknowledgements}
I would like to extend particular thanks to Pat Osmer and the Ohio State
University Astronomy department for their very generous help and hospitality
during the writing of this paper, and for the use of their computing
facilities without which aid this research could not have gone
forward. Thanks are also due to Kip Thorne and Julia Kennefick
for much help and encouragement. Special thanks, for much helpful
and friendly advice, and many interesting conversations go to
Scott Hughes (in particular for his advice on the calculation of
the spheroidal harmonics), 
Eric Poisson, Hideyuki Tagoshi, Masaru Shibata, Amos
Ori and Sam Finn. Thanks also to Oxford University Astrophysics for
their hospitality during the final revision of the paper.
This reseach has been supported by
NSF grant AST-9417371 and NASA grant NAGW-4268.

\section*{Appendix}
The potential functions $F(r)$ and $U(r)$ of the Sasaki-Nakamura
equation (\ref{sn}) are given in this
appendix.
\begin{equation}
F(r)={\eta_{,r}\over \eta} {\Delta\over r^2+a^2}
\end{equation}
where
\begin{equation}
\eta=c_0+c_1/r+c_r/r^2+c_3/r^3+c_4/r^4
\end{equation}
and
\begin{eqnarray}
c_0&=&-12 i \omega M+\lambda(\lambda+2)-12 a \omega(a\omega-m) \\
c_1&=&8 i a[3a\omega-\lambda(a\omega-m)] \\
c_2&=&-24i a M(a\omega-m)+12a^2[1-2(a\omega-m)^2] \\
c_3&=&24i a^3(a\omega-m)-24 M a^2 \\
c_4&=&12 a^4 .
\end{eqnarray}
\begin{equation}
U(r)={\Delta U_1 \over (r^2+a^2)^2}+G^2+{\Delta G_{,r}\over r^2+a^2}-F G
\end{equation}
where
\begin{eqnarray}
G&=&-{2(r-M)\over r^2+a^2}+{r\Delta\over(r^2+a^2)^2} \\
U_1&=&V+{\Delta^2\over \beta}[(2\alpha+{\beta_{,r}\over\Delta})_{,r}
-{\eta_{,r}\over\eta}(\alpha+{\beta_{,r}\over\Delta})] \\
\alpha&=&-i{K\beta\over\Delta^2}+3i K_{,r}+\lambda+{6\Delta\over r^2}
\\
\beta&=&2 \Delta(-i K+r-M-{2\Delta\over r}).
\end{eqnarray}

\clearpage 
\begin{minipage}[t]{2in} 
\begin{table}
\textwidth=3.0cm
\begin{tabular}{l|l} 
$q$ & $r_{\text{crit}}/M$ \\ \hline
-0.9  & 9.64 \\ 
-0.5  & 8.37 \\
0.0   & 6.68 \\
0.5   & 4.70 \\ 
0.7   & 3.76 \\
0.9   & 2.56 \\
0.95  & 2.03 \\
.99   & 1.47 \\
1.0   & 1.0  \\
\end{tabular} 
\caption[Table]{The position of the critical radius, $r_{\text{crit}}$ 
in units of $M$, for different black hole spins $a$. The parameter
$q=a/M$ is defined here to be negative for retrograde orbits and
positive for prograde orbits.}  
\end{table}
\end{minipage} 
\clearpage 
\renewcommand{\baselinestretch}{1}
\large
\normalsize
\begin{figure}
\epsfxsize=\hsize
\epsfbox{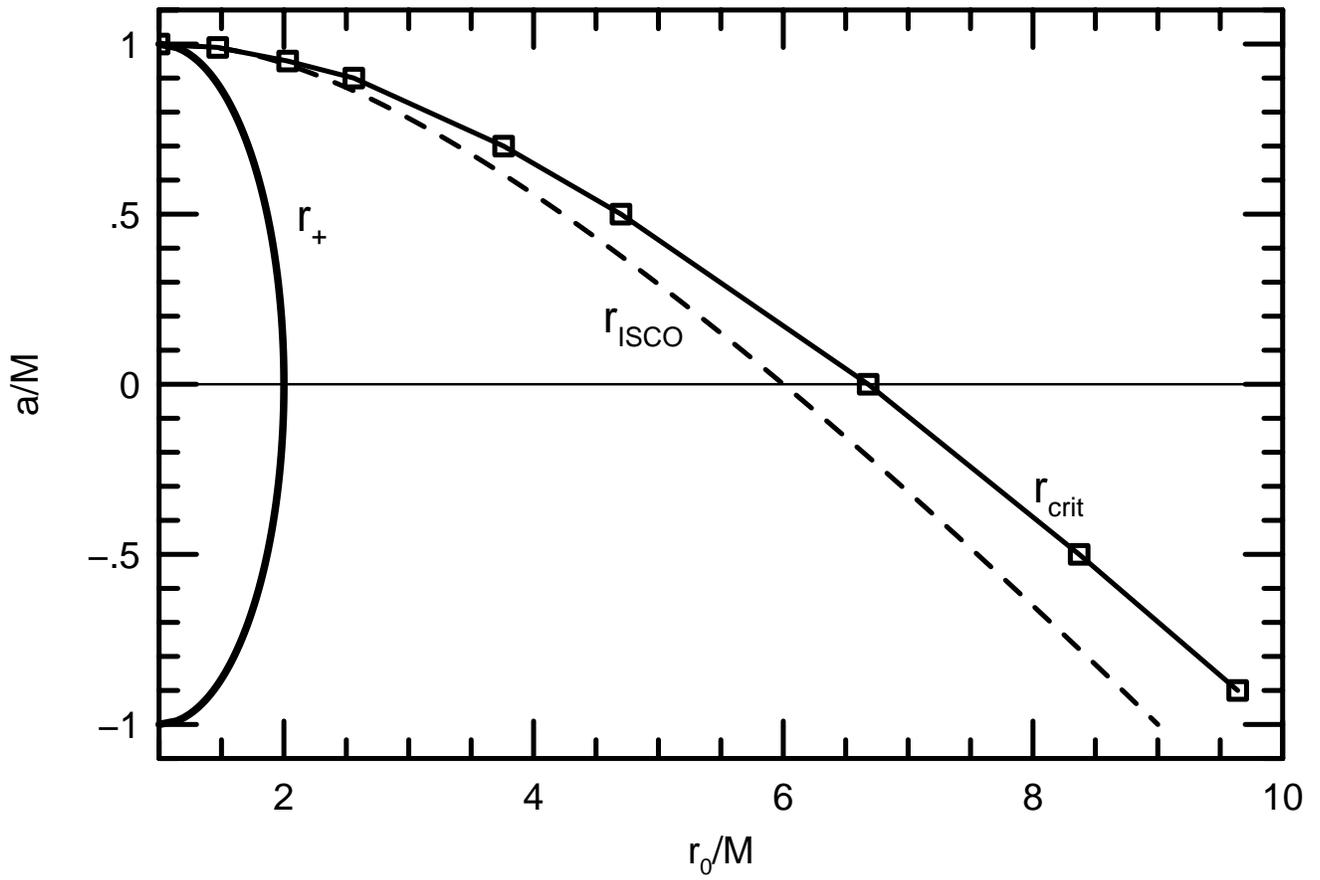}
\caption[Fig. 1]{Graphs showing the positions of the horizon ($r_+$),
innermost stable circular orbit ($r_{\text{ISCO}}$) and critical
radius ($r_{\text{crit}}$) in terms of the mean orbital radius $r_0$ for
all black hole spins ($ a\leq M$). Positive $a$ indicates
a prograde orbit, and negative $a$ a retrograde orbit. The figures 
corresponding to the squares on the critical radius curve (derived
numerically) are given in the accompanying table.}
\end{figure}
\clearpage
\begin{figure}
\epsfxsize=\hsize
\epsfbox{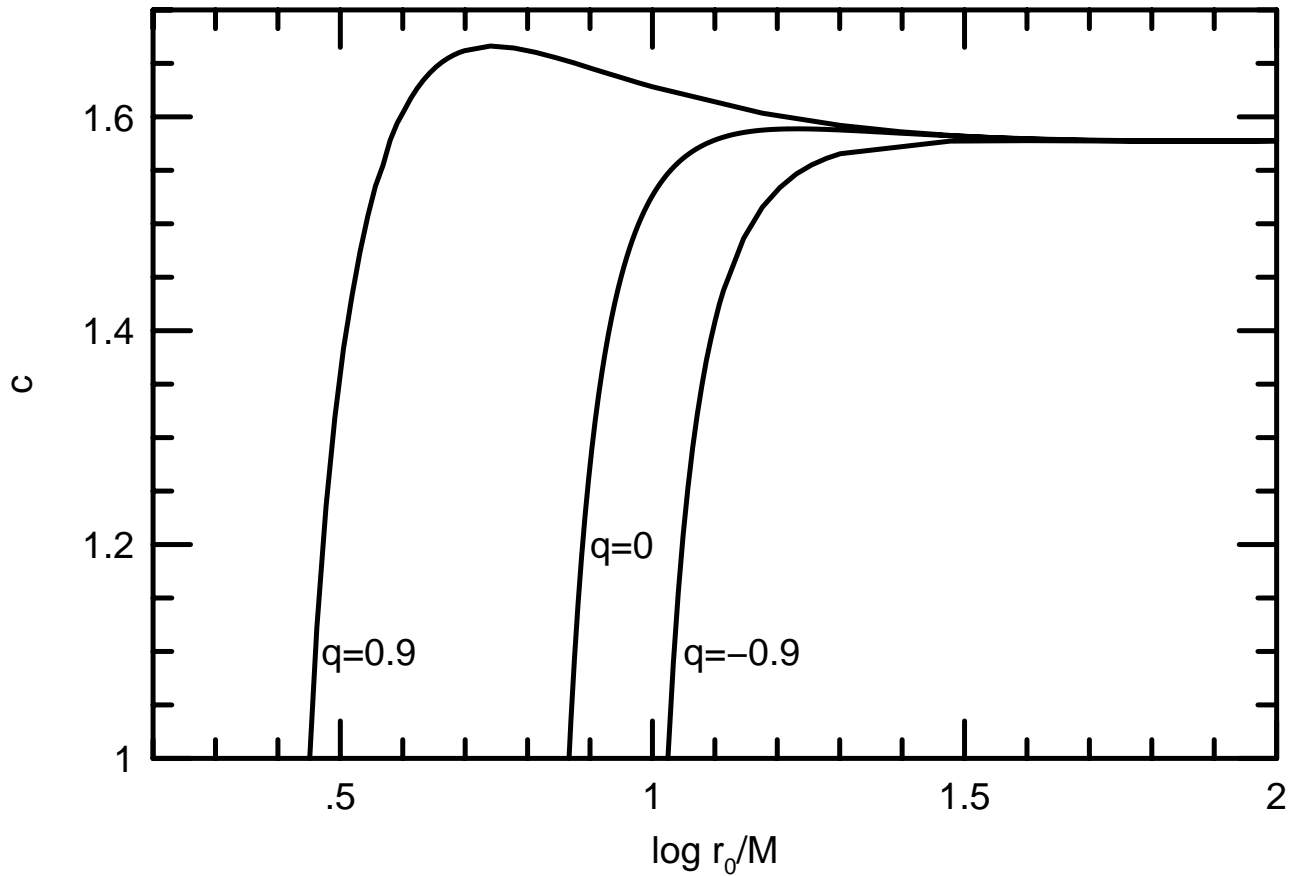}
\caption[Fig. 2]{Curves showing the evolution of the parameter
$c$, defined in Eq.\ (\ref{cee}), as the mean orbital radius
$r_0$ decreases, for three different types of orbit. For
a black hole with spin $a=0.9 M$, both the prograde
($q=a/M=0.9$) and retrograde orbits ($q=-0.9$) are shown. Also
shown is the case of a Schwarzschild black hole ($q=0$).
In each case $c$ begins to fall quickly towards zero as the
innermost stable cirular orbit approaches.}
\end{figure}
\clearpage
\begin{figure}
\epsfxsize=\hsize
\epsfbox{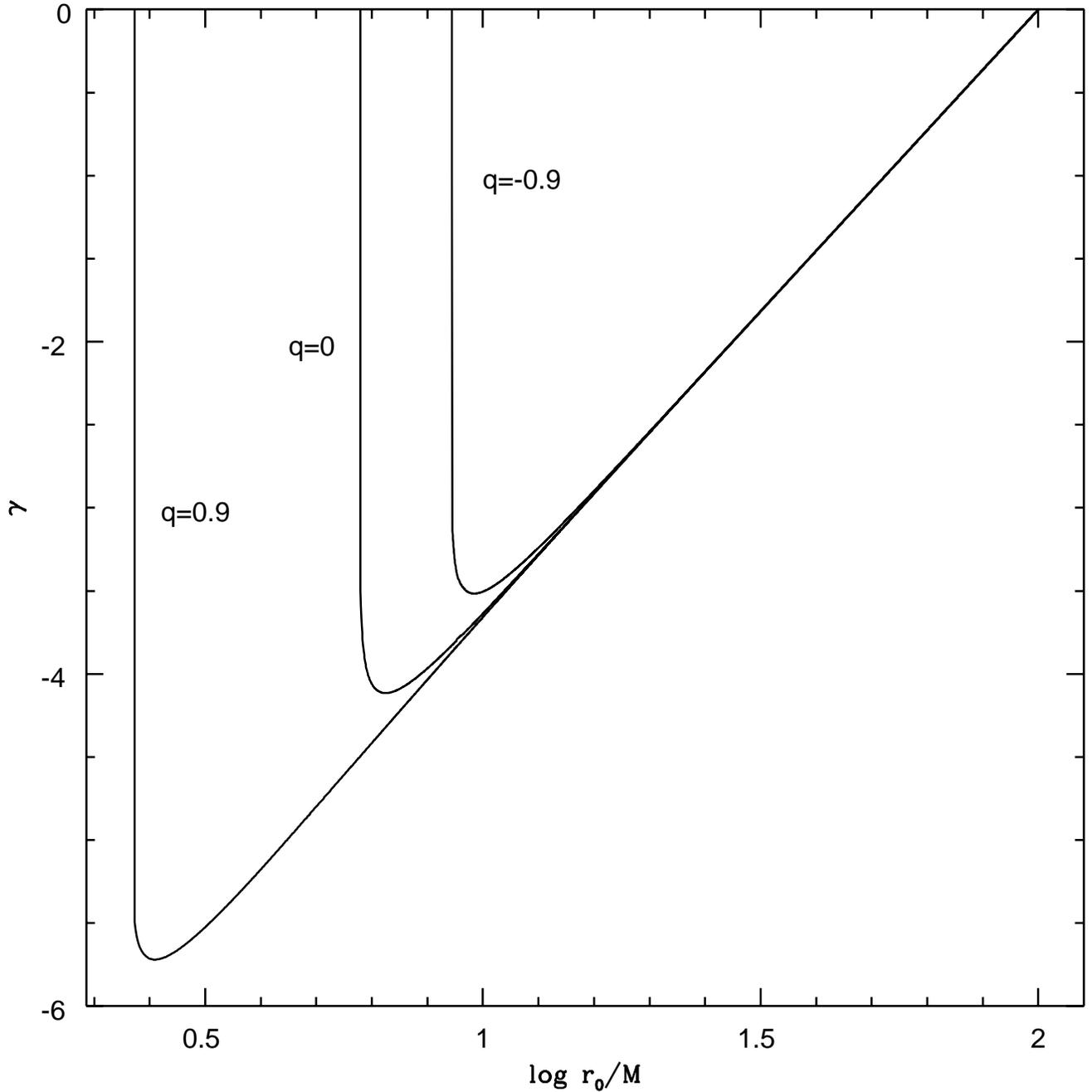}
\caption[Fig. 3]{Curves showing the change of the orbital
eccentricity as the radius $r_0$ decreases, in Schwarzschild
($q=0$), and for prograde ($q=0.9$) and retrograde ($q=-0.9$)
orbits around a Kerr black hole with $a=0.9 M$. In this
graph, the parameter $\gamma$
is the natural log of the ratio of the current eccentricity
(at $r_0$) 
to the eccentricty the orbit had at $r_0=100 M$. We can
see here clearly that at a certain point (equivalent to
the critical radius illustrated in Fig. 1), the eccentricity
begins to increase. For an arbitrarily small mass ratio $\mu/M$
it will increase indefinitely before reaching the ISCO, but
in any practical case this process will be cut off before
too long by the onset of the dynamical instability which causes
the orbiting body to plunge inwards toward the central black hole.}
\end{figure} 
\renewcommand{\baselinestretch}{2}
\large
\normalsize
\large
\end{document}